\documentclass[aps,pre,twocolumn,showcaps,superscriptaddress,showpacs,amsmath,amssymb]{revtex4}

\usepackage{graphicx}
\usepackage{dcolumn}
\usepackage{bm}
\usepackage{color}
\usepackage{upgreek}
\usepackage{color}
\usepackage{placeins}
\usepackage{dsfont}

\newcommand{\eps}{\epsilon}
\newcommand{\al}{\alpha}
\newcommand{\lam}{\lambda}

\renewcommand{\vec}[1]{\mathbf #1}
\newcommand{\mat}[1]{\mathsf #1}
\newcommand{\mean}[1]{\left\langle #1 \right\rangle}

\newcommand{\kB}{\ensuremath{k_{\rm{B}}}}

\renewcommand{\v}{v}
\newcommand{\vecnabla}{\boldsymbol \nabla}

\newcommand{\ex}{\vec{e}_x}

\newcommand{\shr}{\dot{\gamma}}

\newcommand{\nois}{\xi}
\newcommand{\noisvec}{\boldsymbol \nois}
\newcommand{\nb}{N_{\rm b}}
\newcommand{\nbbar}{N_{\rm b}^{\prime}}
\newcommand{\rb}{R_{\rm b}}
\newcommand{\tcryst}{\tau_{\rm x}}
\renewcommand{\r}{\vec{r}}
\renewcommand{\v}{\vec{v}}
\newcommand{\lc}{n}

\newcommand{\dt}{\Delta t}

\newcommand{\csol}{\bar c_{\rm{sol}}}
\newcommand{\cpre}{\bar c_{\rm pre}}
\newcommand{\chcp}{\bar c_{\rm hcp}}
\newcommand{\cbcc}{\bar c_{\rm bcc}}
\newcommand{\cfcc}{\bar c_{\rm fcc}}
\newcommand{\flux}{f}

\newcommand{\T}{T}

\newcommand{\nbonds}{n_{\mathrm{bonds}}}

\newcommand{\wsol}{w^{(0)}_{\mathrm{sol}}}

\newcommand{\whcp}{w^{(0)}_{\mathrm{hcp}}}
\newcommand{\wbcc}{w^{(0)}_{\mathrm{bcc}}}
\newcommand{\wfcc}{w^{(0)}_{\mathrm{fcc}}}
\newcommand{\wvec}{\vec{w}^{(0)}}
\newcommand{\Z}{Z_n}
\newcommand{\one}{\mathds{1}}

\newcommand{\qbarl}{\bar{q}_l}
\newcommand{\qbarsix}{\bar{q}_6}

\newcommand{\wbarl}{\bar{w}_l}
\newcommand{\wbarsix}{\bar{w}_6}
\newcommand{\wbarfour}{\bar{w}_4}
\newcommand{\qbarlm}{\bar{q}_{lm}}
\newcommand{\qbar}{\bar{q}}
\newcommand{\qlm}{q_{lm}}

\newcommand{\wbar}{\bar{w}}

\begin{document}

\title{Crystallization in a sheared colloidal suspension}

\author{Boris Lander}
\author{Udo Seifert}
\affiliation{II. Institut f\"ur Theoretische Physik, Universit\"at Stuttgart,
  Pfaffenwaldring 57, 70550 Stuttgart, Germany}
\author{Thomas Speck}
\affiliation{Institut f\"ur Theoretische Physik II,
  Heinrich-Heine-Universit\"at D\"usseldorf, Universit\"atsstra\ss e 1, 40225
  D\"usseldorf, Germany}

% ---------- Abstract ----------

\begin{abstract}
  We study numerically the crystallization process in a supersaturated
  suspension of repulsive colloidal particles driven by simple shear flow. The
  effect of the shear flow on crystallization is two-fold: while it suppresses
  the initial nucleation, once a large enough critical nucleus has formed its
  growth is enhanced by the shear flow. Combining both effects implies an
  optimal strain rate at which the overall crystallization rate has a
  maximum. To gain insight into the underlying mechanisms, we employ a
  discrete state model describing the transitions between the local structural
  configurations around single particles. We observe a time-scale separation
  between these transitions and the overall progress of the crystallization
  allowing for an effective Markovian description. By using this model, we
  demonstrate that the suppression of nucleation is due to the inhibition of a
  pre-structured liquid.
\end{abstract}


\maketitle

\section{Introduction}

The freezing of a disordered colloidal or nano suspension into a crystalline
state with long-range order is a first-order phase transition that typically
follows the nucleation and growth scenario~\cite{lowe94}. While the stable
crystalline phase is favored energetically, the interface with the disordered
phase penalizes small nuclei and leads to a free energy barrier. Hence, for
crystallization to commence, one has to wait for a rare fluctuation that
results in a large enough \emph{critical nucleus}, which subsequently grows
until it spans the system. This implies that by avoiding nucleation the
density of a suspension can be increased beyond the thermodynamic freezing
density. Such a metastable state is called \emph{supersaturated}. Simple
expressions for the free energy barrier can be obtained within classical
nucleation theory (CNT)~\cite{beck35,frenkel47,das}, which assumes a spherical
nucleus.

Subjecting a colloidal suspension to shear flow drives the system out of
thermal equilibrium. Consequently, the concept of a free-energy is no longer
well-defined and, strictly speaking, CNT cannot be applied
anymore. Nevertheless, sheared colloidal suspensions still crystallize but
with nucleation and growth kinetics that may be significantly altered from
those in equilibrium~\cite{onuk97,verm05}. One might even find a dynamical
coexistence of liquid and solid phases~\cite{butl02,butl03}. Previous studies
of the effects of shear flow on the crystallization rate are not conclusive:
On one side, shear-enhanced crystallization has been reported for
experiments~\cite{acke88,yan94,haw98,amos00,pani02} and numerical
simulations~\cite{moks08,niko11}. On the other side, a suppression of
nucleation has been observed experimentally~\cite{palb95} and
numerically~\cite{butl95,blaa04}. Others report an optimal strain rate for
crystallization in supersaturated hard-sphere-like
suspensions~\cite{holm05,wu09} and protein solutions~\cite{penk06}.  An
optimal strain rate has also been reported in two dimensions for numerical
simulations of Yukawa-type~\cite{cerd08}, Ising~\cite{alle08}, and
depletion-driven attractive~\cite{cerd08} systems, and in a three-dimensional
model glass~\cite{moks10}.

Colloidal systems have the advantage that both the spatial and temporal
evolution can be monitored directly in experiments via light
scattering~\cite{scha93,harl97,sinn01}, or in real space via confocal
microscopy~\cite{gass01,wu09}, see also Ref.~\citenum{gass09} for a
review. Moreover, the interaction potential between colloidal particles can be
tuned~\cite{yeth03}. In the limit of perfect hard-spheres, the crystallization
process is driven solely by entropy and the phase diagram depends on the
density only. For repulsive interactions, temperature plays a role; the
density is, however, still the dominant control parameter. For a numerical
treatment, molecular dynamics as well as Brownian dynamics simulations appear
to be ideally suited to track the temporal development of crystalline nuclei
since particle coordinates are accessible and arbitrary interaction potentials
can be used. However, since nucleation is hampered by a large free-energy
barrier, it is a rare event. Depending on the height of the barrier, sampling
such events can range from difficult to prohibitive unless specifically
tailored methods are employed. Such methods include transition path
sampling~\cite{dell02,pan04}, forward flux sampling~\cite{alle05,alle09}, and
umbrella sampling~\cite{torr77,frenkel}.

In this work, we present numerical results for a model colloidal suspension in
three dimensions, where particles interact via hard-core exclusion plus a
short-ranged Yukawa repulsion. The suspension is strongly supersaturated such
that the probability for nucleation is shifted into a regime accessible by
straightforward methods without making use of importance sampling
schemes. Here, we employ underdamped Langevin dynamics. Although the
suspension is strongly supersaturated, we observe that crystallization still
proceeds via nucleation and growth. For the densities studied, small clusters
appear and disappear until a cluster reaches the critical size. The following
growth process is dominated by a single large cluster. We find that the time
for this largest cluster to reach a specified size (larger than the critical
size but smaller than the total number of particles) decreases for small
strain rates but strongly increases at larger strain rates. There is,
therefore, an optimal strain rate for the crystallization process. The reason
is that shear flow is responsible for two competing effects: suppression of
the initial formation of a critical nucleus and enhancement of its growth once
it has formed. Strain rates considered here are so low that shear-induced
layering plays no role~\cite{acke88,rast96}.

For unsheared liquids it has recently been emphasized that nucleation
resembles a two-stage process and that pre-structuring of the liquid plays a
crucial role in the formation of the critical
nucleus~\cite{luts06,schi10a,lech11,russ12}: Droplets of the stable phase
appear preferentially in regions of the supersaturated liquid that are still
amorphous but where particle have already developed a loose connectivity with
their neighbors. We confirm the two-stage scenario for the model studied here
and demonstrate that the main effect of the shear flow is to disrupt the
formation of such a pre-structured liquid. Hence, we find that the suppressed
nucleation under shear flow has its origin in the inhibition of structuring in
the liquid rather than in the ``demolition'' of crystalline clusters. The
enhanced growth rate of sufficiently large clusters can be attributed to
either convection or a faster reorganization of the cluster, making it thus
easier to incorporate new particles.

The paper is organized as follows. In Sec.~\ref{sec:meth}, we present the
system and simulation details. Moreover, we describe the structural order
parameters that we employ to determine the structure of the local environment
of single particles. In Sec.~\ref{sec:res}, we present and discuss our
simulation results for both the nucleation and growth stage before we conclude
in Sec.~\ref{sec:conc}.

\section{Methods}
\label{sec:meth}

% ----- system -----
\subsection{System and simulation details}
\label{sec:sim}

We consider a mono-disperse colloidal suspension consisting of $N=4860$
particles in a simulation box of constant volume $V$ with volume fraction
$\phi \equiv \pi N/(6V)$. We study two densities with corresponding box
dimensions $17.709\times17.351\times18.404$ for $\phi=0.45$ and
$16.876\times16.535\times17.538$ for $\phi=52$. Throughout the manuscript, we
measure length in units of the particle diameter $a$ and energy in units of
the thermal energy $\kB T$, where $T$ is the temperature of the suspension and
$\kB$ is Boltzmann's constant. The time it takes for a particle to diffuse a
distance corresponding to its diameter $a$ defines the unit of time $3\pi\eta
a^3 / (\kB T)$, where $\eta$ is the viscosity of the solvent.

We employ underdamped Langevin dynamics given by $\dot{\r}_k = \v_k$ and
\begin{equation}
  \label{eq:lang}
  m\dot{\v}_k = -\vecnabla_k U - [\v_k - \vec{u}(\r_k)] + \noisvec_k,
\end{equation}
with positions $\r_k$ and velocities $\v_k$. We choose the dimensionless mass
$m=1$ to be unity such that the relaxation time of the momenta equals the
diffusive time-scale. The interaction forces are described by the potential
\begin{equation}
  U = \sum\limits_{i < j} u(|\vec{r}_i - \vec{r}_j|).
\end{equation}
Thermal fluctuations are modeled by the stochastic forces $\noisvec$ with zero
mean and correlations
\begin{equation}
  \mean{\noisvec_i(t) \noisvec_j^T(0)} = 2\one\delta_{ij}\delta(t),
\end{equation}
where $\delta_{ij}$ denotes the Kronecker symbol, $\delta(t)$ the Dirac
distribution, and $\one$ the identity matrix. Moreover, we impose an external
linear shear flow by means of a solvent velocity field $\vec{u}(\vec{r}) =
\shr y \ex$ entering the friction term, where $\shr$ is the strain rate and
$\ex$ the unit vector in $x$ direction. In the simulations, we employ periodic
boundary conditions using the Lees-Edwards sliding bricks method~\cite{allen}.

Particles interact pairwise via the repulsive Yukawa potential
\begin{equation}
  \label{eq:yuk}
  u(r) =
  \begin{cases}
    \eps\frac{e^{-\kappa(r-1)}}{r} & (r\geqslant 1) \\
    \infty & (r<1)
  \end{cases}
\end{equation}
with hard-core exclusion. The strength of the potential is given by the energy
at contact $\eps$, and its range is determined by the inverse screening length
$\kappa$. The magnitude of $\kappa$ is mainly influenced by the ion
concentration in the solvent and interpolates between Coulombic (low $\kappa$,
low ion concentration) and hard-sphere interactions (large $\kappa$, high ion
concentration). In this study, we choose $\eps = 10$ and $\kappa = 8$. For
this set of parameters, the freezing volume fraction is $\phi^\ast \simeq
0.38$. We are interested in the influence of a weak but steady shear flow on
the crystallization dynamics of colloidal suspensions under highly
supersaturated conditions ($\phi = 0.45$ and $0.52$).

The equations of motion~\eqref{eq:lang} are integrated using a version of the
velocity Verlet algorithm with time step $\dt = 5\times10^{-4}$.  Initial
configurations for different runs are generated by equilibrating the system at
low densities ($\phi = 0.2$). The volume fraction is then increased stepwise
by rescaling the simulation box and the particle coordinates until the final
volume fraction is reached. We use $20000$ time steps. Once we reach the final
volume fraction, we switch on the shear flow with strain rate $\shr$.

% ----- structure analysis -----
\subsection{Structure analysis}

In order to describe the crystallization process quantitatively, we need a way
to distinguish between liquid and solid structures. For the hard-core Yukawa
system, the phase diagram of the equilibrium bulk structure includes, beside
the liquid phase, the two crystalline structures body-centered cubic (bcc) and
face-centered cubic (fcc)~\cite{robb88,meij91,lowe94}. In the limit of
hard-sphere interaction (high $\kappa$), the free-energy difference between an
fcc and a hexagonal close-packed (hcp) configuration is very
low~\cite{wood97}. Therefore, hcp structures are likely to occur along with
fcc structures, as has been observed in microgravity
experiments~\cite{zhu97}. For the parameters studied here, the bulk
equilibrium structure is fcc. However, in the spirit of the Ostwald step
rule~\cite{ostw97}, intermediate structures may be of a different type. Small
nuclei, e.g., have been found to belong predominantly to the bcc structure in
a Lennard-Jones liquid~\cite{wold99}. Hence, in our analysis, we will not only
discern fcc from the liquid state, but also include hcp and bcc-structures.

We employ different variants of the well-known Steinhardt order
parameters~\cite{stei83} to determine the local environment of a single
particle. The basic idea is to construct quantities sensitive to the
rotational symmetry of the local environment of the particles. To that end,
one locates the set of neighbors $\nb(k)$ of the $k$th particle with size
$|\nb(k)|$ and computes the connecting vectors $\vec{r}_{kj} \equiv \vec{r}_k
- \vec{r}_j$ of the central particle with its $j$th neighbor, where a neighbor
is defined as another particle within a range not exceeding $\rb$. This range
is frequently chosen as the minimum between the first and second shell in the
pair correlation function. We define the complex vector
\begin{equation}
  \qlm(k) \equiv \frac{1}{|\nb(k)|} \sum\limits_{j \in \nb(k)} Y_{lm}(\vec{r}_{kj}),
\end{equation}
where the functions $Y_{lm}$ are spherical harmonics and $l>0$, $|m|\leqslant
l$. The vectors $\qlm$ depend sensitively on the choice of $l$.

The first step is two distinguish fluid particles (disordered environment)
from solid particles (ordered environment). To this end, we make use of a
recently introduced variant of the Steinhardt order parameters that averages
over the second-neighbor shell~\cite{lech08},
\begin{equation}
  \qbarl(k) \equiv
  \sqrt{\frac{4\pi}{2l+1}\sum\limits_{m=-1}^{l}\left|\qbarlm(k)\right|},
\end{equation}
where
\begin{equation}
  \qbarlm(k) \equiv \frac{1}{|\nbbar(k)|}\sum\limits_{j \in \nbbar(k)} \qlm(j),
\end{equation}
where $\nbbar(k)$ is the set of neighboring particles including the $k$th
particle itself.  Averaging the order parameter this way sharpens the
distinction between different structures at the expense of spatial
resolution. For $l=6$, the probability distributions of $\qbar_6$ for fluid
and solid particles are well separated, providing a good way to discriminate
these two basic structure types from each other (data not shown).  We regard a
particle as fluid if $\qbarsix<0.4$ and as solid otherwise.

Having determined these two particle sets, we further split the fluid
particles into two subsets: liquid and pre-structured. While liquid particles
have a truly disordered environment, we identify particles as pre-structured
that have an environment that does not qualify as solid but where nevertheless
some ``bonds'' between particles have formed. To concretize the concept of a
bond, we consider the normalized scalar product
\begin{equation}
  \label{eq:S}
  S^{(l)}_{kj} \equiv \frac{\sum\limits_{m = -l}^{l} \qlm(k)
    q^{\ast}_{lm}(j)}{\left(\sum\limits_{m=-l}^l
      \qlm(k)q^{\ast}_{lm}(k)\right)^{1/2} \left(\sum\limits_{m=-l}^l
      \qlm(j)q^{\ast}_{lm}(j)\right)^{1/2}},
\end{equation}
with $q^{\ast}$ the complex conjugate of $q$. This product defines a measure
for the strength of the correlation between the surrounding structures of the
$k$th and the $j$th particle. We regard two neighboring particles as
``bonded'' if $S^{(6)}_{kj}>0.5$~\cite{wold96} and denote the number of bonds
for the $k$th particle as $\nbonds$. While also in the liquid
particles will have bonds, we consider particles that have at least 
$8$ bonds with neighboring particles (but still $\qbarsix<0.4$) as pre-structured.

\begin{figure}
  \centering
  \includegraphics{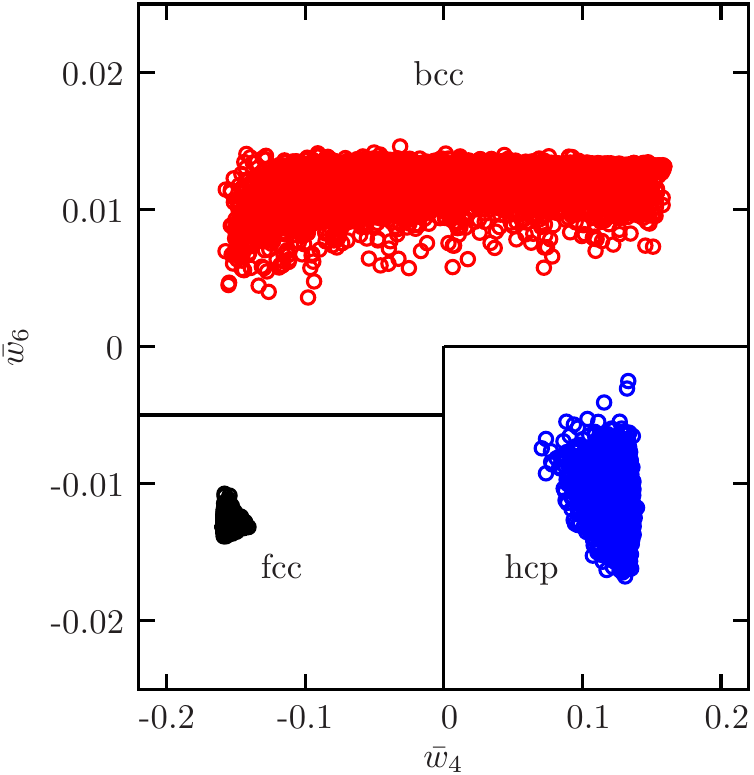}
  \caption{Scatter plot of $\wbarfour$-$\wbarsix$ trajectories for perfect
    hcp, bcc, and fcc crystals subject to thermal fluctuations. The 
    $\wbarfour$-$\wbarsix$ plane is divided into the indicated regions, which 
    are used to determine the crystal structure of solid particles.}
  \label{fig:gauge_w4_w6}
\end{figure}

\begin{figure}
  \centering
  \includegraphics[scale=.7]{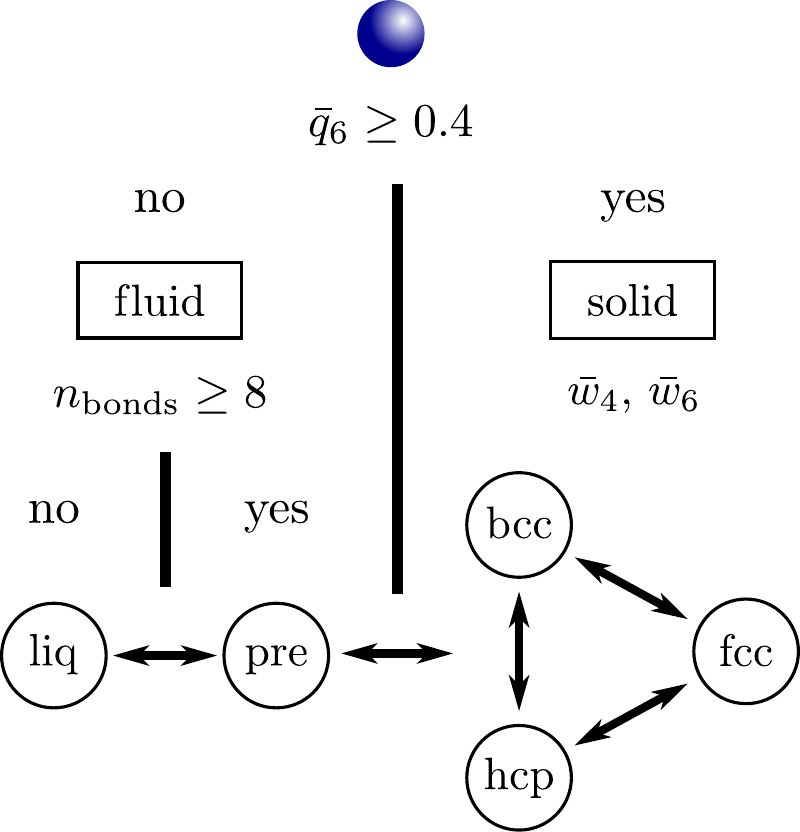}
  \caption{Decision tree for the assignment of a structure to a particle. On
    the first level, the particle is classified either as fluid or solid. On
    the second level, the fluid particles are split into pre-structured and
    liquid particles, while the solid particles are subdivided into the
    crystalline structures hcp, bcc, and fcc. The double-headed arrows
    indicate the transitions we observe primarily during the crystallization
    process: While among the solid structures all transitions occur, the
    liquid particles almost never reach a crystalline state without passing
    through a pre-structured configuration first.}
  \label{fig:struc_assign}
\end{figure}

Finally, by employing another type of averaged order parameters~\cite{lech08}
\begin{equation}
  \wbarl(k) \equiv \frac{\sum\limits_{m_1 + m_2 + m_3 = 0}
    \left( \begin{array}{ccc} 
		  l & l & l\\ 
		  m_1 & m_2 & m_3 
	    \end{array} \right)
    \qbar_{lm_1}(k)\qbar_{lm_2}(k)\qbar_{lm_3}(k)}{\left(
\sum\limits_{m=-l}^{l}|\qbar_{lm}(k)|^2\right)^{3/2}},
\end{equation}
we are able to discern the different crystalline structures within the solid
particles. The term in brackets is the Wigner-3-j symbol, which is related to
Clebsch-Gordan coefficients. The sum runs over all combinations of $m_1$,
$m_2$ and $m_3$ with $m_1+m_2+m_3=0$. Using two parameters $\wbar_4$ and
$\wbar_6$, we obtain a good separation between the distributions in the
$\wbar_4\wbar_6$ plane, see scatter plot in Fig.~\ref{fig:gauge_w4_w6}. The
$\wbarfour$ distribution is widely separated for hcp and fcc structures, while
$\wbarsix$ separates bcc from hcp and fcc. Hence, a solid particle is
classified as fcc if $\wbarfour \le 0$ and $\wbarsix \le -0.005$ (lower left
region), as hcp if $\wbarfour > 0$ and $\wbarsix \le 0$ (lower right region),
and as bcc otherwise (upper region). Fig.~\ref{fig:struc_assign} illustrates
the complete classification process. Moreover, it shows the possible
transitions between the different structures occurring during the
crystallization. Between the solid states, each of the transitions occurs.
However, the transitions are biased towards an fcc environment. As we will see
in the following, direct transitions from the liquid into a crystalline state
are rare and the solidification of a liquid particle generally advances
through a pre-structured state first.

\subsection{Discrete state model}

The structure analysis enables us to categorize single particles according to
their local environment. For a given configuration of particle positions at
time $t$, we define an indicator function $h_k(t)$ for every particle $k$
which takes on one of the five values: liquid (liq), pre-structured (pre),
hcp, bcc, or fcc. The population (fraction of particles) of structure type $i$
is
\begin{equation}
  c_i(t) \equiv \frac{1}{N} \sum_{k=1}^N \delta_{i,h_k(t)}.
\end{equation}
As the suspension evolves, the structural environment of particles will of
course change. To quantify these changes we define the fluxes
\begin{equation}
  \label{eq:fluxes}
  \flux_{i\to j}(t) \equiv \sum\limits_{k=1}^N \delta_{i,h_k(t)}
  \delta_{j,h_k(t+\delta t)},
\end{equation}
which count the number of particles that have been converted from structure
$i$ into structure $j$ within the time interval $[t,t+\delta t]$. If not
indicated otherwise, we set $\delta t = 100 \dt$.

In the following, we consider the size $n$ of the largest cluster as an order
parameter characterizing the progress of the crystallization process. Clusters
are identified as the sets of all solid particles that are mutually bonded (in
the sense defined above that $S^{(6)}_{kj}>0.5$). We define the average
population at fixed cluster size $n$ in the suspension as the conditional
average
\begin{equation}
  \bar c_i(n) \equiv \mean{c_i(t)\delta_{n,n(t)}} / \Z,
\end{equation}
where $\mean{\cdot}$ averages over time and over different realizations 
of the crystallization process. Hence, $\Z \equiv \mean{\delta_{n,n(t)}}$ 
counts how many times the largest cluster size $n$ occurs in all runs 
considered. Moreover, we define a $5\times 5$ transfer matrix $\mat{T}$, 
the components of which are given by the fluxes as
\begin{equation}
  \label{eq:T}
  \T_{i\to j}(n) \equiv \left.\mean{ \frac{\flux_{i\to j}(t)}{N c_i(t)} \delta_{n,n(t)} } 
\right/ \Z
\end{equation}
with normalization $\sum_j\T_{i\to j} = 1$. The component $\T_{i\to j}(n)$ of
this stochastic matrix quantifies the fraction of particles in state $i$ that
convert on average into state $j$ within the time interval $\delta t$. The
eigenvalues of $\mat T$ can be sorted, $\lam_0>\lam_1>\cdots>\lam_4$, with
$\lam_0=1$. The components of the corresponding right-hand-side eigenvectors
fulfill
\begin{equation}
  \sum\limits_{i} w^{(0)}_i = 1, \quad
  \sum\limits_{i} w^{(\al)}_i = 0 \quad (\al \ge 1).
\end{equation}
The product $\mat{T}(n(t)){\vec c}(t)={\vec c}(t+\delta t)$ yields the
average population a time $\delta t$ later with ${\vec c}(t)\equiv(c_i(t))$. 
Applying the transfer matrix repeatedly describes an effective Markovian 
dynamics \emph{at fixed cluster size}. Under this dynamics, the
average population after a time $\tau$ has elapsed becomes
\begin{equation}
  \vec c(\tau) = \vec w^{(0)}+\sum_{\al=1}^4\zeta_\al\vec
  w^{(\al)} e^{-\tau/\tau_\al}
\end{equation}
with implied time scales $\tau_\al(n)\equiv-\delta t/\ln\lam_\al(n)$ and
coefficients $\zeta_\al\equiv\vec w^{(\al)}\cdot \vec c(0)$. Hence, for
$\tau\gg\tau_1$, the system approaches a (quasi)-stationary average population
given by $\wvec$. The relaxation time is determined by $\tau_1$.

\begin{figure}
  \centering
  \includegraphics{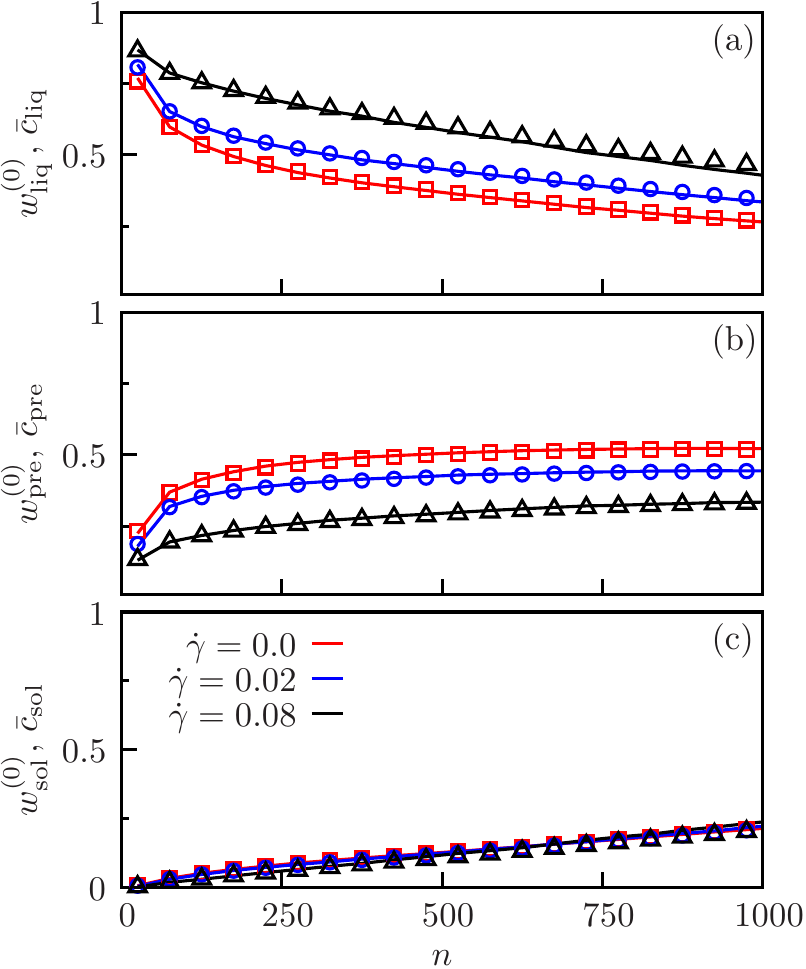}
  \caption{(a)~Liquid, (b)~pre-structured, and (c)~solid fraction of the
    quasi-stationary populations $\wvec$ as function of $n$ (lines) and the
    corresponding actual populations $\bar{\vec c}$ (symbols).
    \label{fig:struc}}
\end{figure}

For a freely evolving suspension, the size of the largest cluster is of course
not constrained to a fixed $n$. Suppose there is a time-scale separation:
After a change of $n$ the system relaxes into the quasi-stationary state
before the cluster size changes again. Then the actual average populations of
local structures measured in the simulations should be equal to the stationary
eigenvector: $\bar{\vec c}(n)\approx\wvec(n)$. Indeed, as shown in
Fig.~\ref{fig:struc}, this is the case to a very good degree. Shown are the
actual and quasi-stationary fractions of particles in a liquid,
pre-structured, and solid local environment, where
$\csol\equiv\chcp+\cbcc+\cfcc$ and $\wsol \equiv \whcp + \wbcc + \wfcc$ sum
the contributions of all crystalline particles. Deviations are small but
increase with increasing strain rate/cluster size up to maximal $7\%$ at
$n=1000$. This demonstrates that the growth of the cluster is a slow process
and that the lag time $\delta t$ is sufficient to sample the fast dynamics.

\section{Results}
\label{sec:res}

% ---- average crystallization rate ----
\subsection{Crystallization rate}

\begin{figure}
 \centering
  \includegraphics{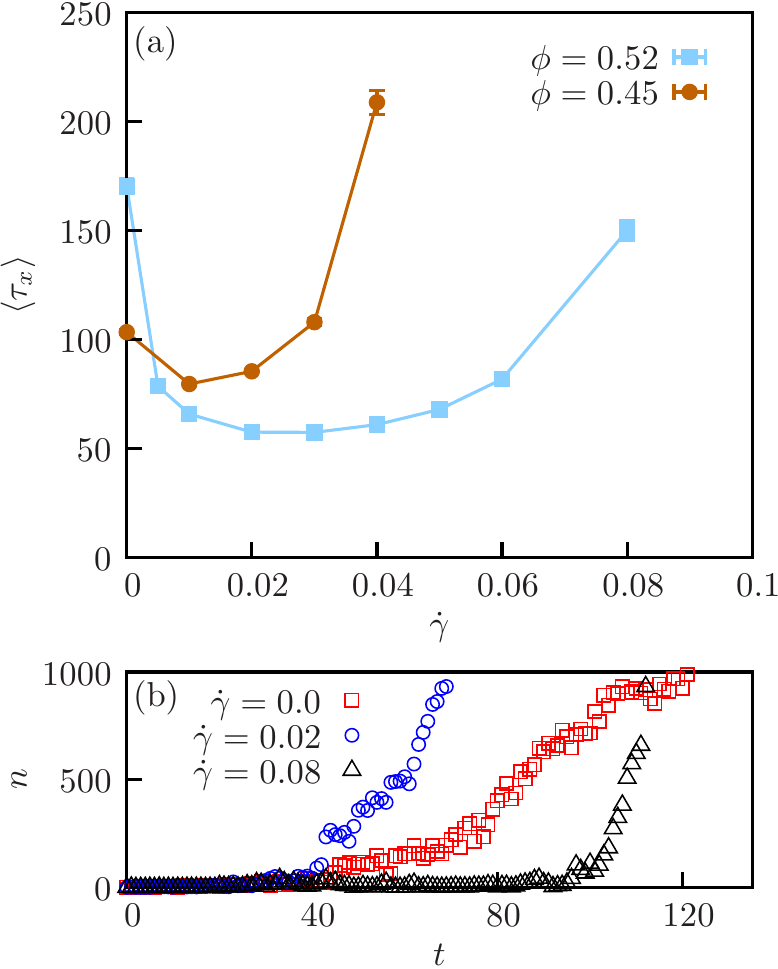}
  \caption{(a)~Average times of crystallization $\tcryst$ as a function of the
    shear rate with standard error bars for the estimation of the mean value. 
    (b)~Temporal development of $n$ for typical runs at $\phi = 0.52$ for the 
    strain rates $\shr = 0.0$, $0.02$, and $0.08$. \label{fig:CrystTimes}}
\end{figure}

We first examine the overall effect of shear flow on the total time it takes a
supersaturated suspension to crystallize. According to the protocol described
above, we start with a supersaturated, fluid suspension and apply shear flow
at $t=0$. Once the largest cluster in the suspension has reached a size of $n
= 800$, we stop the simulation and denote the elapsed time as $\tcryst$. The
stopping size is clearly larger than the critical nucleus size but small
enough (compared to the total number of particles) to minimize finite-size
effects due to the periodic boundaries. Averaging over $500$ independent runs
for each strain rate, we obtain a non-monotonous dependence of $\tcryst$ on
$\shr$, see Fig.~\ref{fig:CrystTimes}(a). For vanishing strain rate, the
denser suspension takes longer to crystallize. Increasing the strain rate,
$\tcryst$ decreases for both densities. For the larger density, however, the
decrease is much stronger and $\tcryst$ drops even below the value for the
less dense system. In both cases, we observe an optimal strain rate
$\shr^{\ast}$ at which the crystallization process is fastest with
$\shr^{\ast} \simeq 0.02$ and $\shr^{\ast} \simeq 0.01$ for $\phi = 0.52$ and
$\phi = 0.45$, respectively. At high strain rates, crystallization becomes
rare.  Already for $\shr = 0.1$ and $\shr = 0.05$, for $\phi = 0.52$ and $\phi
= 0.45$, respectively, a significant part of the $500$ runs did not
crystallize within $2\times10^{6}$ time steps. We thus find an accelerated
crystallization for small but nonvanishing strain rates, while for higher
$\shr$ crystallization is more and more suppressed.

Following the development of the size $n$ of the largest cluster in the
suspension in time, we find qualitative differences in the way the crystalline
state is reached for different strain rates, see Fig.~\ref{fig:CrystTimes}(b).
Without shear flow, the system starts crystallizing quickly but the cluster
grows slowly, whereas in the high-shear case, the system stays at a low $n$
for some time before crystallization is initiated. Afterwards, the largest
cluster develops rapidly. Close to the optimal strain rate, we find features
of both limiting cases. The crystallization process starts almost as early as
without shear flow, but progresses more rapidly later on.

% ---- initial stage ----
\subsection{Shear flow suppresses nucleation}

\subsubsection{Growth rate}

\begin{figure}
 \centering
  \includegraphics{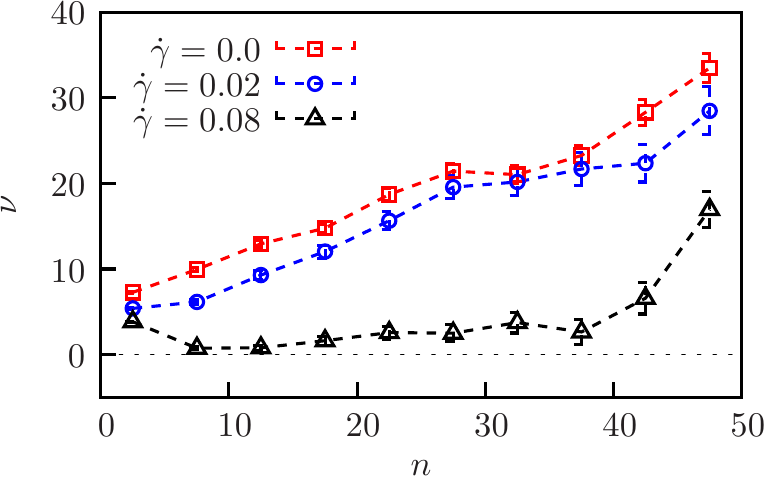}
  \caption{Average growth rate of the size $n$ of the largest cluster at the
    nucleation stage ($n$ small) as a function of $n$ for different strain
    rates at $\phi = 0.52$. For large $\shr$, the growth rate of the largest
    cluster is strongly suppressed in the interval $10 \lesssim n \lesssim
    40$.}
    \label{fig:n_growth_vs_n_small}
\end{figure}

A pertinent quantity to study is the growth rate of the largest cluster
\begin{equation}
  \label{eq:rate:n}
  \nu(n) \equiv \left.\mean{\frac{n(t+\delta t)-n(t)}{\delta t}\delta_{n,n(t)}} \right/ \Z
\end{equation}
averaged at fixed $n$, i.~e., at a specific stage in the development of the
largest cluster. The dot denotes the rate of change. In  
Fig.~\ref{fig:n_growth_vs_n_small}, we show this quantity for $\lc \le 50$. 
For $\shr = 0.02$, the average growth rate is only marginally smaller than 
the one in the unsheared case. For $\shr = 0.08$, however, the growth 
is strongly suppressed in a broad range $10 \lesssim n \lesssim 40$.
This result could be due to a shear-induced inhibition of the formation, or
due to a shear-induced destruction of small clusters. Either way, the 
likelihood for the formation of a critical cluster is strongly reduced. In
the remainder of this subsection, we discuss the relevance of these 
mechanisms. 

\subsubsection{Effect on pre-structured liquid}

Following the two-stage scenario for nucleation, a crystalline cluster is not
likely to occur in the middle of an entirely random distribution of
particles. Rather, in a region of the liquid which has already acquired a
loosely ordered state, fluctuations transforming parts of this pre-structured
liquid into a crystal are much more likely to occur. This scenario also holds
in the system studied here: While the fraction of pre-structured particles
transferring to a solid state is on the order of a few percent, $\T_{\rm pre
  \to sol} \sim 0.03$, the corresponding fraction of particles converting
directly from liquid to solid is smaller by more than three orders of
magnitude, $\T_{\rm liq \to sol} < 10^{-5}$.

\begin{figure}
 \centering
  \includegraphics{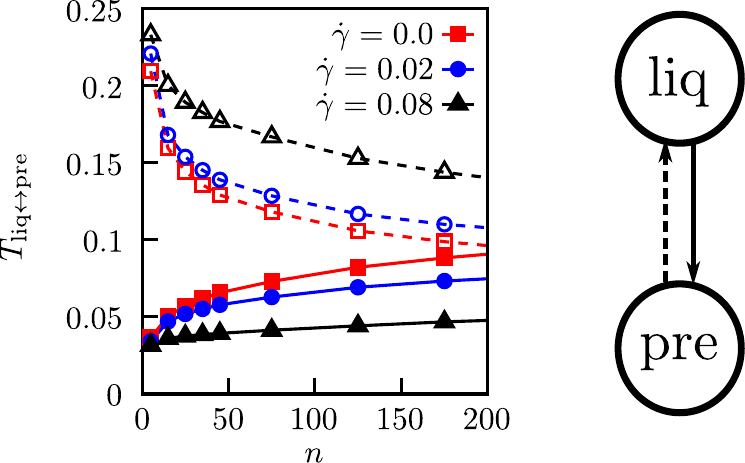}
  \caption{Transfer matrix components for transitions from the liquid to the
    pre-structured state (solid lines) and the corresponding backwards
    transitions (dashed lines) for volume fraction $\phi = 0.52$ as a function
    of cluster size $n$.
    \label{fig:liq2struc}}
\end{figure}

In order to study the effect of shear flow on the structure of the liquid, we
record the transfer matrix components $\T_{\rm liq \to pre}$ and corresponding 
backward component $\T_{\rm pre \to liq}$, see Fig.~\ref{fig:liq2struc}. Note
that $\T_{i \to j}$ describes the \emph{fraction} of particles in structure $i$ 
transferring on average to structure $j$ in the following time interval $\delta t$. 
Hence, the actual net current of particles changing from $i$ to $j$ depends 
as well on the population in these states. Hence, although $\T_{\rm liq \to pre} < 
\T_{\rm pre \to liq}$, the net current of particles is still directed from liquid 
to pre-structured, as the liquid state contains much more particles than the 
pre-structured one at this early stage in the crystallization process.
In Fig.~\ref{fig:liq2struc}, we show $\T_{\rm liq \to pre}$ and 
$\T_{\rm pre \to liq}$ for three different strain rates. We find that shear flow 
has a significant influence on the development of structure in the liquid. On 
the one hand, the establishment of bonds is inhibited, as can be seen from 
the reduced values for $\T_{\rm liq \to pre}$. On the other hand, structure in 
the liquid is destroyed, resulting in an enhanced value for $\T_{\rm pre \to liq}$. 
Consequently, compared to the unsheared case, we find a much smaller 
fraction of pre-structured particles $\cpre$ both in the stationary and in the 
actual composition of the suspension for $\shr = 0.08$, see 
Fig.~\ref{fig:struc}(b). In other words, shear flow prevents the liquid from 
developing a loose structure.

\begin{figure}
 \centering
  \includegraphics{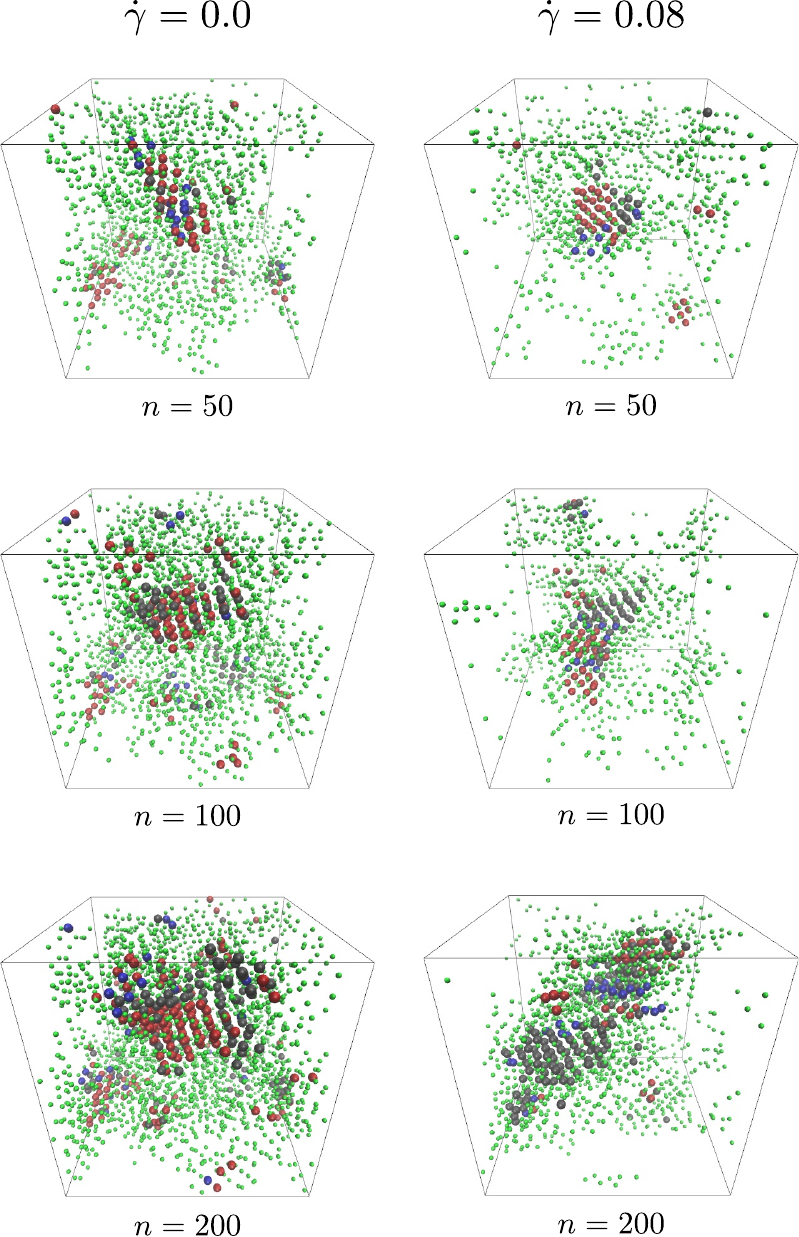}
  \caption{Snapshots of the suspension during the crystallization process for
    $\shr = 0.0$ (left column) and $\shr = 0.08$ (right column) for three
    different sizes of the largest cluster $n$.  The large spheres are solid
    particles with hcp (blue), bcc (gray), and fcc-structured (red)
    environments. Pre-structured particles are shown as small sphere (green),
    whereas liquid particles are not shown for clarity.}
  \label{fig:snapshots}
\end{figure}

In Fig.~\ref{fig:snapshots}, we follow the evolution of the largest
crystalline cluster for strain rates $\shr = 0.0$ and $0.08$. Clearly, in both
cases the crystallization process is dominated by a single cluster. Note that
crystalline clusters (large spheres in blue, gray, and red) are composed of
different local structures, but there seems to be no tendency for a certain
structure type to form a core or surface (e.g., a core of fcc particles with a
cluster-fluid interface formed by bcc particles). Crystalline clusters are
surrounded by pre-structured particle (small green spheres). Although most
prominent in the vicinity of solid clusters, this loose structure can be found
throughout the suspension. In the strongly sheared case, however, we observe
considerably less pre-structured particles than in the unsheared suspension,
which indicates the shear-induced disruption of a loosely structured fluid.

\subsubsection{From pre-structured to solid}

\begin{figure}
 \centering
  \includegraphics[width=1.0\linewidth]{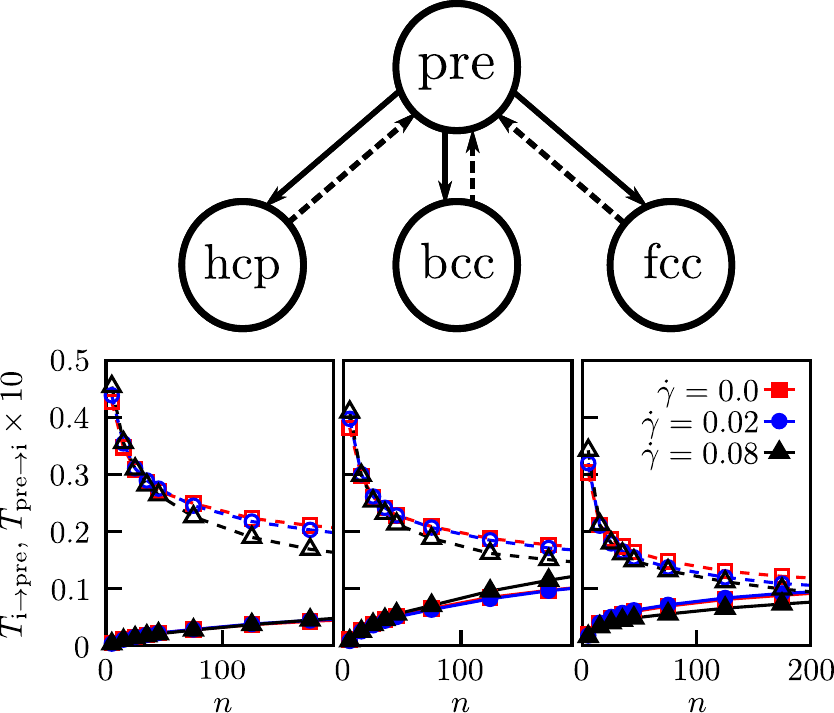}
  \caption{Transfer matrix components for transitions from the pre-structured
    state to hcp, bcc, and fcc (from left to right, solid lines) scaled by a
    factor $10$ and their corresponding backwards fluxes (dashed) for $\shr =
    0.0$, $0.02$, and $0.08$. \label{fig:struc2sol}}
\end{figure}

Once a pre-structured but still amorphous environment has formed, the
pre-structured liquid has yet to transform into a crystalline cluster. Hence,
the next step is to focus on the influence of shear flow on the second part of
the transition from liquid to crystal. We trace the transitions between
pre-structured and the crystalline structures hcp, bcc, and fcc and show the
corresponding transfer components $\T_{j\to i}$ in
Fig.~\ref{fig:struc2sol}. Here, the influence of the shear flow is much
smaller than for the transitions between liquid and pre-structured. The rates
from the crystalline states to the pre-structured one describe the destruction
of crystalline clusters. Interestingly, these are not enhanced by the shear
flow but even somewhat reduced. Transitions into the different crystalline
states are also barely affected by the shear flow. Hence, we find that the
shear flow is not strong enough to actually destroy or shrink clusters once
they have formed.

% ---- growth stage ----
\subsection{Shear flow enhances growth of clusters}

\subsubsection{Growth rate}

\begin{figure}
  \centering
  \includegraphics{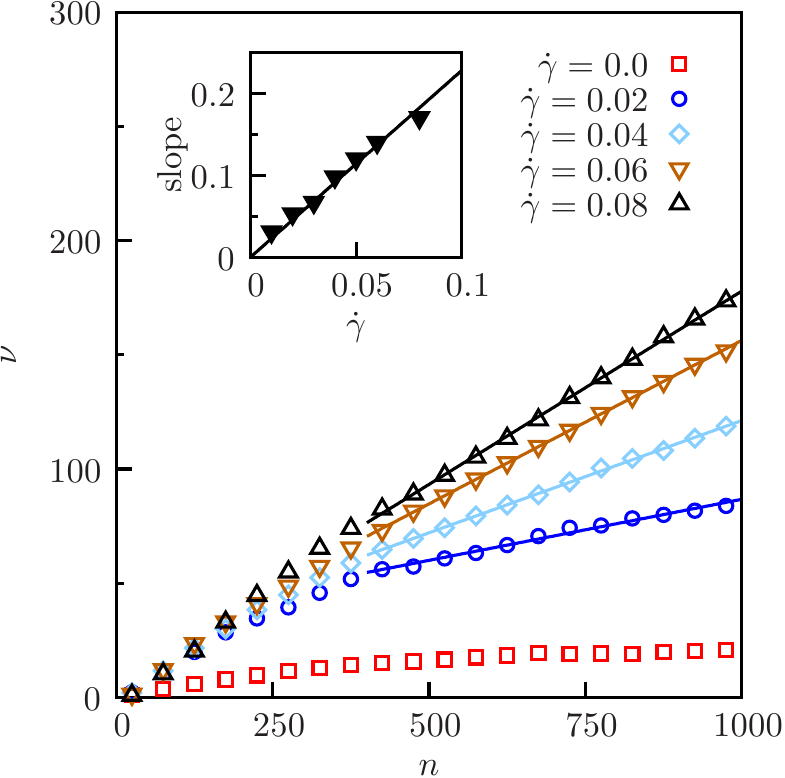}
  \caption{Average growth rate of the largest cluster as a function of $n$ for
    different strain rates for volume fraction $\phi = 0.52$. The solid lines
    are linear fits to the data for $n \ge 400$. Inset: Slope of the fit
    functions as a function of strain rate $\shr$. The solid line is a linear
    fit starting at the origin.}
  \label{fig:lc_growth_vs_lc}
\end{figure}

Complementary to Fig.~\ref{fig:n_growth_vs_n_small}, we now
plot the growth rate Eq.~\eqref{eq:rate:n} over a wider range of 
cluster sizes shown in Fig.~\ref{fig:lc_growth_vs_lc}. Note that 
for the larger clusters considered here, we need to increase the 
time interval over which the change in $n$ is evaluated to 
$\delta t = 5000 \dt$ in order to separate the growth trend from 
the fluctuations. Compared to the unsheared case, the growth 
rate is enhanced significantly once the shear flow is switched on. 
Furthermore, for cluster sizes $n\gtrsim400$ the growth rate is 
dominated by a linear term, the slopes of which themselves grow 
proportional with the strain rate, see inset in 
Fig.~\ref{fig:lc_growth_vs_lc}. We observe the same behavior for 
$\phi = 0.45$ (data not shown). The dominant contribution of the 
shear flow to the growth rate for $n \gtrsim 400$ is thus a linear 
term of the form
\begin{equation}
  \label{eq:conv_contribution}
  \nu(n) = B \shr n,
\end{equation}
where the proportionality constants for the slopes $B$ can be determined from
least-square fits to the data. We find $B = 2.28 \pm 0.10$ and $2.75\pm 0.16$
for $\phi = 0.52$ and $\phi = 0.45$, respectively. This strongly
shear-dependent growth rate reflects the enhanced cluster growth for large
times as shown in Fig.~\ref{fig:CrystTimes}(b) on a more systematic level.  
We now discuss two possible mechanisms of how shear flow influences the 
growth stage of crystallization.

\subsubsection{Mechanism I: Convection}

A growth rate that is a linear function of $n$ might arise from
convection, which can be understood as follows. We assume 
that particles in the vicinity of a crystalline nucleus are more likely 
to crystallize. Shear flow enhances the number of particles passing 
through this direct vicinity of the nucleus which we model as a sphere 
with radius $R$. The particle current entering this sphere caused by 
the shear flow reads
\begin{equation}
  \frac{1}{2} \int_{\mathcal{S}} \rho_l \shr|y \ex \cdot d\vec{S}| 
  = \frac{4}{3}\rho_{l}R^3 \shr \propto \shr n
\end{equation}
with the surface of the sphere $\mathcal{S}$ and the number density 
of particles in the liquid $\rho_{l} \equiv \pi\phi/6$. The particle current 
is thus proportional to the size of the cluster, and proportional to the 
strain rate $\shr$. Hence, we obtain the functional form of 
Eq.~\eqref{eq:conv_contribution} where the free parameter $B$ takes 
into account deviations from the spherical shapes of the cluster and 
the probability with which particles become solid (attach to the cluster).

\subsubsection{Mechanism II: bcc grows fastest}

\begin{figure}
  \centering
  \includegraphics{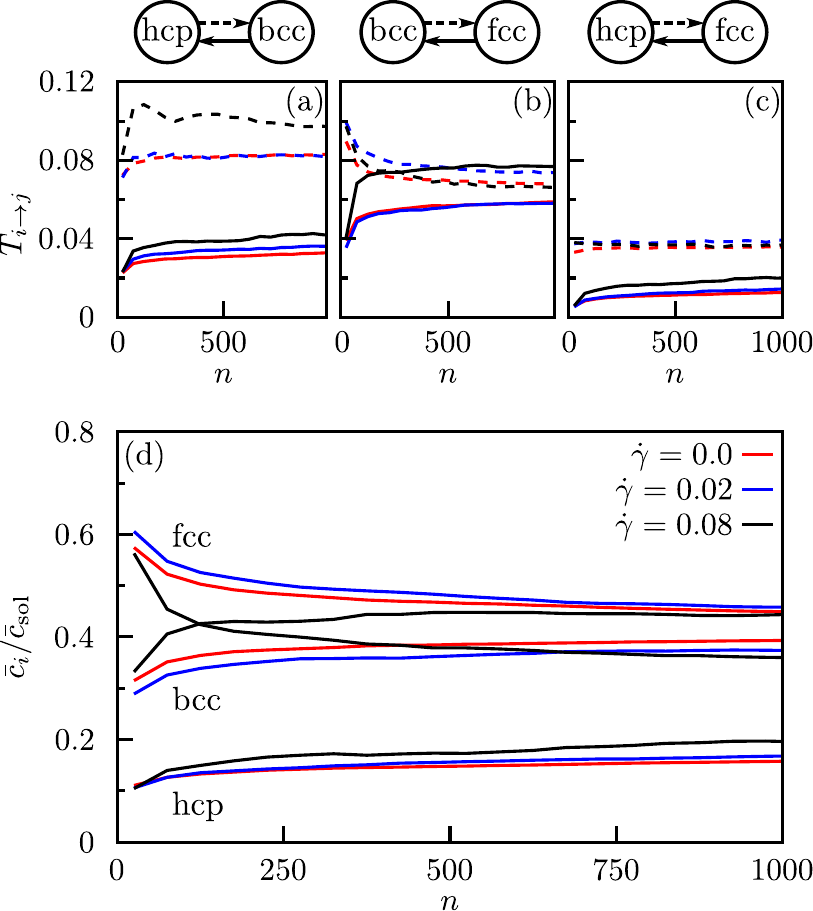}
  \caption{Transfer matrix components for the transitions between the three
    solid local structures: (a)~hcp and bcc, (b)~bcc and fcc, and (c)~hcp and
    fcc. (d)~Relative populations of solid particles in structure hcp, bcc,
    and fcc $\bar{c}_i/ \csol$ measured in the simulations.}
  \label{fig:crystal}
\end{figure}

A second explanation for an enhanced growth rate might be that one of 
the local structures can be grown faster. In Fig.~\ref{fig:crystal}(a)-(c), 
we show the components of the transfer matrix for transitions between 
the three solid structures. The most prominent effect of the shear flow 
is that the largest strain rate strongly facilitates transitions towards the 
bcc structure. For the other transitions, the effect of the shear flow is 
weaker. The rates towards the hcp structure grow slightly with increasing 
$\shr$ and the optimal strain rate $\shr = 0.02$ enhances transitions from 
bcc to fcc.

Fig.~\ref{fig:crystal}(d) shows the relative composition of all solid
particles in the suspension as a function of cluster size $n$. For small $\shr$,
most solid particles belong to fcc, which is the stable equilibrium
structure. Interestingly, even in the unsheared case does the fraction of fcc
particles slightly decrease as the cluster becomes bigger. In consistency with
the shear dependence of the transition rates, at the optimal strain rate
$\shr=0.02$, the fraction of bcc particles is decreased and the number of fcc
particles is increased, whereas the shear flow has almost no influence of the
fraction of hcp particles.  Quite in contrast, the bcc particles overtake the
fcc particles at the higher strain rate $\shr=0.08$ and become the dominant
structural type. As can be seen in the snapshots in Fig.~\ref{fig:snapshots},
shear flow allows for much larger bcc domains. This can be interpreted as a
consequence of the Ostwald step rule: in the unsheared case bcc is more easily
formed but fcc is the more stable structure. Shear flow stabilizes bcc in
relation to fcc and hcp, thus allowing for larger fractions of this structure
type. If bcc is indeed the fastest growing structure, this mechanism would
also lead to an enhanced crystalline growth.

\section{Conclusions}
\label{sec:conc}

We have performed Langevin dynamics simulations to investigate the overall
effect of weak linear shear flow on the crystallization process in a model
colloidal suspension. We found that the time it takes for the suspension to
become solid exhibits a minimum at a finite strain rate. This can be explained
as the result of two counteracting effects: at the early nucleation stage, the
shear flow inhibits the formation of a critical nucleus, while at the later
stage growth is enhanced. Both effects become manifest in the growth rate of
the largest cluster, see Fig.~\ref{fig:n_growth_vs_n_small} and
Fig.~\ref{fig:lc_growth_vs_lc}. For larger strain rates the growth rate of
small clusters is vanishingly small, whereas already small strain rates
significantly speed up the growth process of the crystalline clusters after
nucleation.

To gain further insight into the different mechanism, we have employed a
discrete state model. The state space of this discrete model comprises the
relevant local structural environments of a \emph{single} particle. The
microscopic environment of a particle might be fluid or solid. We further
distinguish between the unordered liquid and a pre-structured liquid, which
still is amorphous but where already a loose structure between neighbors has
formed as determined from a bond criterion. The solid particles are classified
according to their crystal structure as either fcc, bcc, or hcp; leading to a
total of five discrete states. From the simulation data we can measure the
populations, i.e., the fraction of particles that resides in each
structure. Since the simulations advance from the initial liquid towards the
full crystal, these populations change over time. We have found it convenient
to calculate conditional averages with respect to the size of the largest
cluster. Moreover, we have determined the fluxes between the discrete states,
from which we built a transfer matrix. The quasi-stationary state of this
transfer matrix agrees well with the measured actual populations.

Nucleation in our model system clearly proceeds from liquid through
pre-structured to one of the crystalline structures. The effect of shear flow
is to disrupt the formation of the pre-structured regions, see
Fig.~\ref{fig:liq2struc}. Consequently, there are fewer pre-structured
particles and the probability to form a critical nucleus is drastically
lowered. On the other hand, once a critical droplet of solid particles has
formed, the growth rate is enhanced by the shear flow. The emerging functional
form~\eqref{eq:conv_contribution} can be explained by convection alone: The
flow field constantly changes the vicinity of crystalline clusters and is thus
able to enhance the number of liquid particles under the ``influence'' of a
crystalline cluster. However, the shear flow also influences the composition
of the cluster, hinting at a second mechanism. We found that shear flow
facilitates transitions towards the bcc structure and thus enhances the
fraction of bcc particles in the cluster. The stable bulk structure is fcc
and, therefore, there is a driving force towards fcc. Destabilizing fcc (or
stabilizing bcc) allows for larger domains of bcc particles without the need
to convert bcc, which in turn allows the cluster to grow faster.

Our data show that the depth of the minimum in the duration for the
crystallization process is much smaller for a lower supersaturation. This
point entails the question whether there is a lower bound on the density below
which this minimum vanishes. How the existence and depth of this minimum
depends on the supersaturation remains a topic for future
investigation. Moreover, it would be interesting to study the influence of
hydrodynamic interactions on the crystallization process~\cite{radu13} under
shear in general, and on the magnitude and the existence of such an optimal
strain rate in particular. Beyond the insights we could gain for our specific
model, we believe that the combination of techniques presented here will prove
useful in the study of nucleation and related processes.

% ---- acknowledgments ----
\acknowledgments

We acknowledge financial support by Deutsche Forschungsgemeinschaft (Grant
No. SE-1119/3).

% ---- bibliography ----


\begin{thebibliography}{52}
\expandafter\ifx\csname natexlab\endcsname\relax\def\natexlab#1{#1}\fi
\expandafter\ifx\csname bibnamefont\endcsname\relax
  \def\bibnamefont#1{#1}\fi
\expandafter\ifx\csname bibfnamefont\endcsname\relax
  \def\bibfnamefont#1{#1}\fi
\expandafter\ifx\csname citenamefont\endcsname\relax
  \def\citenamefont#1{#1}\fi
\expandafter\ifx\csname url\endcsname\relax
  \def\url#1{\texttt{#1}}\fi
\expandafter\ifx\csname urlprefix\endcsname\relax\def\urlprefix{URL }\fi
\providecommand{\bibinfo}[2]{#2}
\providecommand{\eprint}[2][]{\url{#2}}

\bibitem[{\citenamefont{L\"owen}(1994)}]{lowe94}
\bibinfo{author}{\bibfnamefont{H.}~\bibnamefont{L\"owen}},
  \bibinfo{journal}{Phys.\ Rep.} \textbf{\bibinfo{volume}{237}},
  \bibinfo{pages}{249} (\bibinfo{year}{1994}).

\bibitem[{\citenamefont{Becker and D{\"o}ring}(1935)}]{beck35}
\bibinfo{author}{\bibfnamefont{R.}~\bibnamefont{Becker}} \bibnamefont{and}
  \bibinfo{author}{\bibfnamefont{W.}~\bibnamefont{D{\"o}ring}},
  \bibinfo{journal}{Annalen der Physik} \textbf{\bibinfo{volume}{416}},
  \bibinfo{pages}{719} (\bibinfo{year}{1935}).

\bibitem[{\citenamefont{Frenkel}(1947)}]{frenkel47}
\bibinfo{author}{\bibfnamefont{J.}~\bibnamefont{Frenkel}},
  \emph{\bibinfo{title}{Kinetic theory of liquids}}
  (\bibinfo{publisher}{Clarendon Pr.}, \bibinfo{address}{Oxford},
  \bibinfo{year}{1947}).

\bibitem[{\citenamefont{Das}(2011)}]{das}
\bibinfo{author}{\bibfnamefont{S.~P.} \bibnamefont{Das}},
  \emph{\bibinfo{title}{Statistical Physics of Liquids at Freezing and Beyond}}
  (\bibinfo{publisher}{Cambridge Univ. Press}, \bibinfo{year}{2011}).

\bibitem[{\citenamefont{Onuki}(1997)}]{onuk97}
\bibinfo{author}{\bibfnamefont{A.}~\bibnamefont{Onuki}}, \bibinfo{journal}{J.
  Phys.: Condens. Matter} \textbf{\bibinfo{volume}{9}}, \bibinfo{pages}{6119}
  (\bibinfo{year}{1997}).

\bibitem[{\citenamefont{Vermant and Solomon}(2005)}]{verm05}
\bibinfo{author}{\bibfnamefont{J.}~\bibnamefont{Vermant}} \bibnamefont{and}
  \bibinfo{author}{\bibfnamefont{M.~J.} \bibnamefont{Solomon}},
  \bibinfo{journal}{J. Phys.: Condens. Matter} \textbf{\bibinfo{volume}{17}},
  \bibinfo{pages}{R187} (\bibinfo{year}{2005}).

\bibitem[{\citenamefont{Butler and Harrowell}(2002)}]{butl02}
\bibinfo{author}{\bibfnamefont{S.}~\bibnamefont{Butler}} \bibnamefont{and}
  \bibinfo{author}{\bibfnamefont{P.}~\bibnamefont{Harrowell}},
  \bibinfo{journal}{Nature} \textbf{\bibinfo{volume}{415}},
  \bibinfo{pages}{1008} (\bibinfo{year}{2002}).

\bibitem[{\citenamefont{Butler and Harrowell}(2003)}]{butl03}
\bibinfo{author}{\bibfnamefont{S.}~\bibnamefont{Butler}} \bibnamefont{and}
  \bibinfo{author}{\bibfnamefont{P.}~\bibnamefont{Harrowell}},
  \bibinfo{journal}{J. Chem. Phys.} \textbf{\bibinfo{volume}{118}},
  \bibinfo{pages}{4115} (\bibinfo{year}{2003}).

\bibitem[{\citenamefont{Ackerson and Pusey}(1988)}]{acke88}
\bibinfo{author}{\bibfnamefont{B.~J.} \bibnamefont{Ackerson}} \bibnamefont{and}
  \bibinfo{author}{\bibfnamefont{P.~N.} \bibnamefont{Pusey}},
  \bibinfo{journal}{Phys.\ Rev.\ Lett.} \textbf{\bibinfo{volume}{61}},
  \bibinfo{pages}{1033} (\bibinfo{year}{1988}).

\bibitem[{\citenamefont{Yan et~al.}(1994)\citenamefont{Yan, Dhont, Smits, and
  Lekkerkerker}}]{yan94}
\bibinfo{author}{\bibfnamefont{Y.~D.} \bibnamefont{Yan}},
  \bibinfo{author}{\bibfnamefont{J.~K.~G.} \bibnamefont{Dhont}},
  \bibinfo{author}{\bibfnamefont{C.}~\bibnamefont{Smits}}, \bibnamefont{and}
  \bibinfo{author}{\bibfnamefont{H.~N.~W.} \bibnamefont{Lekkerkerker}},
  \bibinfo{journal}{Physica A} \textbf{\bibinfo{volume}{202}},
  \bibinfo{pages}{68} (\bibinfo{year}{1994}).

\bibitem[{\citenamefont{Haw et~al.}(1998)\citenamefont{Haw, Poon, and
  Pusey}}]{haw98}
\bibinfo{author}{\bibfnamefont{M.~D.} \bibnamefont{Haw}},
  \bibinfo{author}{\bibfnamefont{W.~C.~K.} \bibnamefont{Poon}},
  \bibnamefont{and} \bibinfo{author}{\bibfnamefont{P.~N.} \bibnamefont{Pusey}},
  \bibinfo{journal}{Phys.\ Rev.\ E} \textbf{\bibinfo{volume}{57}},
  \bibinfo{pages}{6859} (\bibinfo{year}{1998}).

\bibitem[{\citenamefont{Amos et~al.}(2000)\citenamefont{Amos, Rarity, Tapster,
  Shepherd, and Kitson}}]{amos00}
\bibinfo{author}{\bibfnamefont{R.~M.} \bibnamefont{Amos}},
  \bibinfo{author}{\bibfnamefont{J.~G.} \bibnamefont{Rarity}},
  \bibinfo{author}{\bibfnamefont{P.~R.} \bibnamefont{Tapster}},
  \bibinfo{author}{\bibfnamefont{T.~J.} \bibnamefont{Shepherd}},
  \bibnamefont{and} \bibinfo{author}{\bibfnamefont{S.~C.}
  \bibnamefont{Kitson}}, \bibinfo{journal}{Phys.\ Rev.\ E}
  \textbf{\bibinfo{volume}{61}}, \bibinfo{pages}{2929} (\bibinfo{year}{2000}).

\bibitem[{\citenamefont{Panine et~al.}(2002)\citenamefont{Panine, Naranyan,
  Vermant, and Mewis}}]{pani02}
\bibinfo{author}{\bibfnamefont{P.}~\bibnamefont{Panine}},
  \bibinfo{author}{\bibfnamefont{T.}~\bibnamefont{Naranyan}},
  \bibinfo{author}{\bibfnamefont{J.}~\bibnamefont{Vermant}}, \bibnamefont{and}
  \bibinfo{author}{\bibfnamefont{J.}~\bibnamefont{Mewis}},
  \bibinfo{journal}{Phys.\ Rev.\ E} \textbf{\bibinfo{volume}{66}},
  \bibinfo{pages}{022401} (\bibinfo{year}{2002}).

\bibitem[{\citenamefont{Mokshin and Barrat}(2008)}]{moks08}
\bibinfo{author}{\bibfnamefont{A.~V.} \bibnamefont{Mokshin}} \bibnamefont{and}
  \bibinfo{author}{\bibfnamefont{J.-L.} \bibnamefont{Barrat}},
  \bibinfo{journal}{Phys.\ Rev.\ E} \textbf{\bibinfo{volume}{77}},
  \bibinfo{pages}{021505} (\bibinfo{year}{2008}).

\bibitem[{\citenamefont{Nikoubashman et~al.}(2011)\citenamefont{Nikoubashman,
  Kahl, and Likos}}]{niko11}
\bibinfo{author}{\bibfnamefont{A.}~\bibnamefont{Nikoubashman}},
  \bibinfo{author}{\bibfnamefont{G.}~\bibnamefont{Kahl}}, \bibnamefont{and}
  \bibinfo{author}{\bibfnamefont{C.~N.} \bibnamefont{Likos}},
  \bibinfo{journal}{Phys.\ Rev.\ Lett.} \textbf{\bibinfo{volume}{107}},
  \bibinfo{pages}{068302} (\bibinfo{year}{2011}).

\bibitem[{\citenamefont{Palberg et~al.}(1995)\citenamefont{Palberg, M{\"o}nch,
  Schwarz, and Leiderer}}]{palb95}
\bibinfo{author}{\bibfnamefont{T.}~\bibnamefont{Palberg}},
  \bibinfo{author}{\bibfnamefont{W.}~\bibnamefont{M{\"o}nch}},
  \bibinfo{author}{\bibfnamefont{J.}~\bibnamefont{Schwarz}}, \bibnamefont{and}
  \bibinfo{author}{\bibfnamefont{P.}~\bibnamefont{Leiderer}},
  \bibinfo{journal}{J.\ Chem.\ Phys.} \textbf{\bibinfo{volume}{102}},
  \bibinfo{pages}{5082} (\bibinfo{year}{1995}).

\bibitem[{\citenamefont{Butler and Harrowell}(1995)}]{butl95}
\bibinfo{author}{\bibfnamefont{S.}~\bibnamefont{Butler}} \bibnamefont{and}
  \bibinfo{author}{\bibfnamefont{P.}~\bibnamefont{Harrowell}},
  \bibinfo{journal}{Phys.\ Rev.\ E} \textbf{\bibinfo{volume}{52}},
  \bibinfo{pages}{6} (\bibinfo{year}{1995}).

\bibitem[{\citenamefont{Blaak et~al.}(2004)\citenamefont{Blaak, Auer, Frenkel,
  and L\"owen}}]{blaa04}
\bibinfo{author}{\bibfnamefont{R.}~\bibnamefont{Blaak}},
  \bibinfo{author}{\bibfnamefont{S.}~\bibnamefont{Auer}},
  \bibinfo{author}{\bibfnamefont{D.}~\bibnamefont{Frenkel}}, \bibnamefont{and}
  \bibinfo{author}{\bibfnamefont{H.}~\bibnamefont{L\"owen}},
  \bibinfo{journal}{Phys. Rev. Lett.} \textbf{\bibinfo{volume}{93}}
  (\bibinfo{year}{2004}).

\bibitem[{\citenamefont{Holmqvist et~al.}(2005)\citenamefont{Holmqvist,
  Lettinga, Buitenhuis, and Dhont}}]{holm05}
\bibinfo{author}{\bibfnamefont{P.}~\bibnamefont{Holmqvist}},
  \bibinfo{author}{\bibfnamefont{M.~P.} \bibnamefont{Lettinga}},
  \bibinfo{author}{\bibfnamefont{J.}~\bibnamefont{Buitenhuis}},
  \bibnamefont{and} \bibinfo{author}{\bibfnamefont{J.~K.~G.}
  \bibnamefont{Dhont}}, \bibinfo{journal}{Langmuir}
  \textbf{\bibinfo{volume}{21}}, \bibinfo{pages}{10976} (\bibinfo{year}{2005}).

\bibitem[{\citenamefont{Wu et~al.}(2009)\citenamefont{Wu, Derks, van Blaaderen,
  and Imhof}}]{wu09}
\bibinfo{author}{\bibfnamefont{Y.~L.} \bibnamefont{Wu}},
  \bibinfo{author}{\bibfnamefont{D.}~\bibnamefont{Derks}},
  \bibinfo{author}{\bibfnamefont{A.}~\bibnamefont{van Blaaderen}},
  \bibnamefont{and} \bibinfo{author}{\bibfnamefont{A.}~\bibnamefont{Imhof}},
  \bibinfo{journal}{Proc.\ Natl.\ Acad.\ Sci.\ U.S.A.}
  \textbf{\bibinfo{volume}{106}}, \bibinfo{pages}{10564}
  (\bibinfo{year}{2009}).

\bibitem[{\citenamefont{Penkova et~al.}(2006)\citenamefont{Penkova, Pan,
  Hodjaouglu, and Vekilov}}]{penk06}
\bibinfo{author}{\bibfnamefont{A.}~\bibnamefont{Penkova}},
  \bibinfo{author}{\bibfnamefont{W.}~\bibnamefont{Pan}},
  \bibinfo{author}{\bibfnamefont{F.}~\bibnamefont{Hodjaouglu}},
  \bibnamefont{and} \bibinfo{author}{\bibfnamefont{P.~G.}
  \bibnamefont{Vekilov}}, \bibinfo{journal}{Ann. N. Y. Acad. Sci.}
  \textbf{\bibinfo{volume}{1077}}, \bibinfo{pages}{214} (\bibinfo{year}{2006}).

\bibitem[{\citenamefont{Cerd\`a et~al.}(2008)\citenamefont{Cerd\`a, Sintes,
  Holm, Sorensen, and Chakrabarti}}]{cerd08}
\bibinfo{author}{\bibfnamefont{J.~J.} \bibnamefont{Cerd\`a}},
  \bibinfo{author}{\bibfnamefont{T.}~\bibnamefont{Sintes}},
  \bibinfo{author}{\bibfnamefont{C.}~\bibnamefont{Holm}},
  \bibinfo{author}{\bibfnamefont{C.~M.} \bibnamefont{Sorensen}},
  \bibnamefont{and}
  \bibinfo{author}{\bibfnamefont{A.}~\bibnamefont{Chakrabarti}},
  \bibinfo{journal}{Phys. Rev. E} \textbf{\bibinfo{volume}{78}},
  \bibinfo{pages}{031403} (\bibinfo{year}{2008}).

\bibitem[{\citenamefont{Allen et~al.}(2008)\citenamefont{Allen, Valeriani,
  Tanase-Nicola, ten Wolde, and Frenkel}}]{alle08}
\bibinfo{author}{\bibfnamefont{R.~J.} \bibnamefont{Allen}},
  \bibinfo{author}{\bibfnamefont{C.}~\bibnamefont{Valeriani}},
  \bibinfo{author}{\bibfnamefont{S.}~\bibnamefont{Tanase-Nicola}},
  \bibinfo{author}{\bibfnamefont{P.~R.} \bibnamefont{ten Wolde}},
  \bibnamefont{and} \bibinfo{author}{\bibfnamefont{D.}~\bibnamefont{Frenkel}},
  \bibinfo{journal}{J.\ Chem.\ Phys.} \textbf{\bibinfo{volume}{129}},
  \bibinfo{pages}{134704} (\bibinfo{year}{2008}).

\bibitem[{\citenamefont{Mokshin and Barrat}(2010)}]{moks10}
\bibinfo{author}{\bibfnamefont{A.~V.} \bibnamefont{Mokshin}} \bibnamefont{and}
  \bibinfo{author}{\bibfnamefont{J.-L.} \bibnamefont{Barrat}},
  \bibinfo{journal}{Phys.\ Rev.\ E} \textbf{\bibinfo{volume}{82}},
  \bibinfo{pages}{021505} (\bibinfo{year}{2010}).

\bibitem[{\citenamefont{Sch\"atzel and Ackerson}(1993)}]{scha93}
\bibinfo{author}{\bibfnamefont{K.}~\bibnamefont{Sch\"atzel}} \bibnamefont{and}
  \bibinfo{author}{\bibfnamefont{B.~J.} \bibnamefont{Ackerson}},
  \bibinfo{journal}{Phys. Rev. E} \textbf{\bibinfo{volume}{48}},
  \bibinfo{pages}{3766} (\bibinfo{year}{1993}).

\bibitem[{\citenamefont{Harland and van Megen}(1997)}]{harl97}
\bibinfo{author}{\bibfnamefont{J.~L.} \bibnamefont{Harland}} \bibnamefont{and}
  \bibinfo{author}{\bibfnamefont{W.}~\bibnamefont{van Megen}},
  \bibinfo{journal}{Phys. Rev. E} \textbf{\bibinfo{volume}{55}},
  \bibinfo{pages}{3054} (\bibinfo{year}{1997}).

\bibitem[{\citenamefont{Sinn et~al.}(2001)\citenamefont{Sinn, Heymann, Stipp,
  and Palberg}}]{sinn01}
\bibinfo{author}{\bibfnamefont{C.}~\bibnamefont{Sinn}},
  \bibinfo{author}{\bibfnamefont{A.}~\bibnamefont{Heymann}},
  \bibinfo{author}{\bibfnamefont{A.}~\bibnamefont{Stipp}}, \bibnamefont{and}
  \bibinfo{author}{\bibfnamefont{T.}~\bibnamefont{Palberg}},
  \bibinfo{journal}{Prog. Colloid Polym. Sci.} \textbf{\bibinfo{volume}{118}},
  \bibinfo{pages}{266} (\bibinfo{year}{2001}).

\bibitem[{\citenamefont{Gasser et~al.}(2001)\citenamefont{Gasser, Weeks,
  Schofield, Pusey, and Weitz}}]{gass01}
\bibinfo{author}{\bibfnamefont{U.}~\bibnamefont{Gasser}},
  \bibinfo{author}{\bibfnamefont{E.~R.} \bibnamefont{Weeks}},
  \bibinfo{author}{\bibfnamefont{A.}~\bibnamefont{Schofield}},
  \bibinfo{author}{\bibfnamefont{P.~N.} \bibnamefont{Pusey}}, \bibnamefont{and}
  \bibinfo{author}{\bibfnamefont{D.~A.} \bibnamefont{Weitz}},
  \bibinfo{journal}{Science} \textbf{\bibinfo{volume}{292}},
  \bibinfo{pages}{258} (\bibinfo{year}{2001}).

\bibitem[{\citenamefont{Gasser}(2009)}]{gass09}
\bibinfo{author}{\bibfnamefont{U.}~\bibnamefont{Gasser}}, \bibinfo{journal}{J.
  Phys.: Condens. Matter} \textbf{\bibinfo{volume}{21}},
  \bibinfo{pages}{203101} (\bibinfo{year}{2009}).

\bibitem[{\citenamefont{Yethiraj and van Blaaderen}(2003)}]{yeth03}
\bibinfo{author}{\bibfnamefont{A.}~\bibnamefont{Yethiraj}} \bibnamefont{and}
  \bibinfo{author}{\bibfnamefont{A.}~\bibnamefont{van Blaaderen}},
  \bibinfo{journal}{Nature} \textbf{\bibinfo{volume}{421}},
  \bibinfo{pages}{513} (\bibinfo{year}{2003}).

\bibitem[{\citenamefont{Dellago et~al.}(2002)\citenamefont{Dellago, Bolhuis,
  and Geissler}}]{dell02}
\bibinfo{author}{\bibfnamefont{C.}~\bibnamefont{Dellago}},
  \bibinfo{author}{\bibfnamefont{P.~G.} \bibnamefont{Bolhuis}},
  \bibnamefont{and} \bibinfo{author}{\bibfnamefont{P.~L.}
  \bibnamefont{Geissler}}, \bibinfo{journal}{Adv. Chem. Phys.}
  \textbf{\bibinfo{volume}{123}}, \bibinfo{pages}{1} (\bibinfo{year}{2002}).

\bibitem[{\citenamefont{Pan and Chandler}(2004)}]{pan04}
\bibinfo{author}{\bibfnamefont{A.~C.} \bibnamefont{Pan}} \bibnamefont{and}
  \bibinfo{author}{\bibfnamefont{D.}~\bibnamefont{Chandler}},
  \bibinfo{journal}{J. Phys. Chem. B} \textbf{\bibinfo{volume}{108}},
  \bibinfo{pages}{19681} (\bibinfo{year}{2004}).

\bibitem[{\citenamefont{Allen et~al.}(2005)\citenamefont{Allen, Warren, and ten
  Wolde}}]{alle05}
\bibinfo{author}{\bibfnamefont{R.~J.} \bibnamefont{Allen}},
  \bibinfo{author}{\bibfnamefont{P.~B.} \bibnamefont{Warren}},
  \bibnamefont{and} \bibinfo{author}{\bibfnamefont{P.~R.} \bibnamefont{ten
  Wolde}}, \bibinfo{journal}{Phys.\ Rev.\ Lett.} \textbf{\bibinfo{volume}{94}},
  \bibinfo{pages}{018104} (\bibinfo{year}{2005}).

\bibitem[{\citenamefont{Allen et~al.}(2009)\citenamefont{Allen, Valeriani, and
  ten Wolde}}]{alle09}
\bibinfo{author}{\bibfnamefont{R.~J.} \bibnamefont{Allen}},
  \bibinfo{author}{\bibfnamefont{C.}~\bibnamefont{Valeriani}},
  \bibnamefont{and} \bibinfo{author}{\bibfnamefont{P.~R.} \bibnamefont{ten
  Wolde}}, \bibinfo{journal}{J. Phys.: Condens. Matter}
  \textbf{\bibinfo{volume}{21}}, \bibinfo{pages}{463102}
  (\bibinfo{year}{2009}).

\bibitem[{\citenamefont{Torrie and Valleau}(1977)}]{torr77}
\bibinfo{author}{\bibfnamefont{G.~M.} \bibnamefont{Torrie}} \bibnamefont{and}
  \bibinfo{author}{\bibfnamefont{J.~P.} \bibnamefont{Valleau}},
  \bibinfo{journal}{J.\ Comput.\ Phys.} \textbf{\bibinfo{volume}{23}},
  \bibinfo{pages}{187} (\bibinfo{year}{1977}).

\bibitem[{\citenamefont{Frenkel and Smit}(2002)}]{frenkel}
\bibinfo{author}{\bibfnamefont{D.}~\bibnamefont{Frenkel}} \bibnamefont{and}
  \bibinfo{author}{\bibfnamefont{B.}~\bibnamefont{Smit}},
  \emph{\bibinfo{title}{Understanding Molecular Simulation: From Algorithms to
  Applications}} (\bibinfo{publisher}{Academic Press, San Diego},
  \bibinfo{year}{2002}).

\bibitem[{\citenamefont{Rastogi et~al.}(1996)\citenamefont{Rastogi, Wagner, and
  Lustig}}]{rast96}
\bibinfo{author}{\bibfnamefont{S.~R.} \bibnamefont{Rastogi}},
  \bibinfo{author}{\bibfnamefont{N.~J.} \bibnamefont{Wagner}},
  \bibnamefont{and} \bibinfo{author}{\bibfnamefont{S.~R.}
  \bibnamefont{Lustig}}, \bibinfo{journal}{J.\ Chem.\ Phys.}
  \textbf{\bibinfo{volume}{104}}, \bibinfo{pages}{9234} (\bibinfo{year}{1996}).

\bibitem[{\citenamefont{Lutsko and Nicolis}(2006)}]{luts06}
\bibinfo{author}{\bibfnamefont{J.~F.} \bibnamefont{Lutsko}} \bibnamefont{and}
  \bibinfo{author}{\bibfnamefont{G.}~\bibnamefont{Nicolis}},
  \bibinfo{journal}{Phys. Rev. Lett.} \textbf{\bibinfo{volume}{96}},
  \bibinfo{pages}{046102} (\bibinfo{year}{2006}).

\bibitem[{\citenamefont{Schilling et~al.}(2010)\citenamefont{Schilling,
  Sch\"ope, Oettel, Opletal, and Snook}}]{schi10a}
\bibinfo{author}{\bibfnamefont{T.}~\bibnamefont{Schilling}},
  \bibinfo{author}{\bibfnamefont{H.~J.} \bibnamefont{Sch\"ope}},
  \bibinfo{author}{\bibfnamefont{M.}~\bibnamefont{Oettel}},
  \bibinfo{author}{\bibfnamefont{G.}~\bibnamefont{Opletal}}, \bibnamefont{and}
  \bibinfo{author}{\bibfnamefont{I.}~\bibnamefont{Snook}},
  \bibinfo{journal}{Phys. Rev. Lett.} \textbf{\bibinfo{volume}{105}},
  \bibinfo{pages}{025701} (\bibinfo{year}{2010}).

\bibitem[{\citenamefont{Lechner et~al.}(2011)\citenamefont{Lechner, Dellago,
  and Bolhuis}}]{lech11}
\bibinfo{author}{\bibfnamefont{W.}~\bibnamefont{Lechner}},
  \bibinfo{author}{\bibfnamefont{C.}~\bibnamefont{Dellago}}, \bibnamefont{and}
  \bibinfo{author}{\bibfnamefont{P.~G.} \bibnamefont{Bolhuis}},
  \bibinfo{journal}{Phys. Rev. Lett.} \textbf{\bibinfo{volume}{106}},
  \bibinfo{pages}{085701} (\bibinfo{year}{2011}).

\bibitem[{\citenamefont{Russo and Tanaka}(2012)}]{russ12}
\bibinfo{author}{\bibfnamefont{J.}~\bibnamefont{Russo}} \bibnamefont{and}
  \bibinfo{author}{\bibfnamefont{H.}~\bibnamefont{Tanaka}},
  \bibinfo{journal}{Sci. Rep.} \textbf{\bibinfo{volume}{2}}
  (\bibinfo{year}{2012}).

\bibitem[{\citenamefont{Allen and Tildesley}(1987)}]{allen}
\bibinfo{author}{\bibfnamefont{M.~P.} \bibnamefont{Allen}} \bibnamefont{and}
  \bibinfo{author}{\bibfnamefont{D.~J.} \bibnamefont{Tildesley}},
  \emph{\bibinfo{title}{Computer Simulation of Liquids}}
  (\bibinfo{publisher}{Clarendon Press}, \bibinfo{address}{Oxford},
  \bibinfo{year}{1987}).

\bibitem[{\citenamefont{Robbins et~al.}(1988)\citenamefont{Robbins, Kremer, and
  Grest}}]{robb88}
\bibinfo{author}{\bibfnamefont{M.~O.} \bibnamefont{Robbins}},
  \bibinfo{author}{\bibfnamefont{K.}~\bibnamefont{Kremer}}, \bibnamefont{and}
  \bibinfo{author}{\bibfnamefont{G.~S.} \bibnamefont{Grest}},
  \bibinfo{journal}{J.\ Chem.\ Phys.} \textbf{\bibinfo{volume}{88}},
  \bibinfo{pages}{3286} (\bibinfo{year}{1988}).

\bibitem[{\citenamefont{Meijer and Frenkel}(1991)}]{meij91}
\bibinfo{author}{\bibfnamefont{E.~J.} \bibnamefont{Meijer}} \bibnamefont{and}
  \bibinfo{author}{\bibfnamefont{D.}~\bibnamefont{Frenkel}},
  \bibinfo{journal}{J.\ Chem.\ Phys.} \textbf{\bibinfo{volume}{94}},
  \bibinfo{pages}{2269} (\bibinfo{year}{1991}).

\bibitem[{\citenamefont{Woodcock}(1997)}]{wood97}
\bibinfo{author}{\bibfnamefont{L.~V.} \bibnamefont{Woodcock}},
  \bibinfo{journal}{Nature} \textbf{\bibinfo{volume}{385}},
  \bibinfo{pages}{141} (\bibinfo{year}{1997}).

\bibitem[{\citenamefont{Zhu et~al.}(1997)\citenamefont{Zhu, Li, Rogers, Meyer,
  Ottewill, Crew, Russel, and Chaikin}}]{zhu97}
\bibinfo{author}{\bibfnamefont{J.}~\bibnamefont{Zhu}},
  \bibinfo{author}{\bibfnamefont{M.}~\bibnamefont{Li}},
  \bibinfo{author}{\bibfnamefont{R.}~\bibnamefont{Rogers}},
  \bibinfo{author}{\bibfnamefont{W.}~\bibnamefont{Meyer}},
  \bibinfo{author}{\bibfnamefont{R.~H.} \bibnamefont{Ottewill}},
  \bibinfo{author}{\bibfnamefont{S.-. S.~S.} \bibnamefont{Crew}},
  \bibinfo{author}{\bibfnamefont{W.~B.} \bibnamefont{Russel}},
  \bibnamefont{and} \bibinfo{author}{\bibfnamefont{P.~M.}
  \bibnamefont{Chaikin}}, \bibinfo{journal}{Nature}
  \textbf{\bibinfo{volume}{387}}, \bibinfo{pages}{883} (\bibinfo{year}{1997}).

\bibitem[{\citenamefont{Ostwald}(1897)}]{ostw97}
\bibinfo{author}{\bibfnamefont{W.}~\bibnamefont{Ostwald}}, \bibinfo{journal}{Z.
  Phys. Chem.} \textbf{\bibinfo{volume}{22}}, \bibinfo{pages}{289}
  (\bibinfo{year}{1897}).

\bibitem[{\citenamefont{ten Wolde and Frenkel}(1999)}]{wold99}
\bibinfo{author}{\bibfnamefont{P.~R.} \bibnamefont{ten Wolde}}
  \bibnamefont{and} \bibinfo{author}{\bibfnamefont{D.}~\bibnamefont{Frenkel}},
  \bibinfo{journal}{Phys. Chem. Chem. Phys.} \textbf{\bibinfo{volume}{1}},
  \bibinfo{pages}{2191} (\bibinfo{year}{1999}).

\bibitem[{\citenamefont{Steinhardt et~al.}(1983)\citenamefont{Steinhardt,
  Nelson, and Ronchetti}}]{stei83}
\bibinfo{author}{\bibfnamefont{P.~J.} \bibnamefont{Steinhardt}},
  \bibinfo{author}{\bibfnamefont{D.~R.} \bibnamefont{Nelson}},
  \bibnamefont{and}
  \bibinfo{author}{\bibfnamefont{M.}~\bibnamefont{Ronchetti}},
  \bibinfo{journal}{Phys.\ Rev.\ B} \textbf{\bibinfo{volume}{28}},
  \bibinfo{pages}{2} (\bibinfo{year}{1983}).

\bibitem[{\citenamefont{Lechner and Dellago}(2008)}]{lech08}
\bibinfo{author}{\bibfnamefont{W.}~\bibnamefont{Lechner}} \bibnamefont{and}
  \bibinfo{author}{\bibfnamefont{C.}~\bibnamefont{Dellago}},
  \bibinfo{journal}{J.\ Chem.\ Phys.} \textbf{\bibinfo{volume}{129}},
  \bibinfo{pages}{114707} (\bibinfo{year}{2008}).

\bibitem[{\citenamefont{ten Wolde and Ruiz-Montero}(1996)}]{wold96}
\bibinfo{author}{\bibfnamefont{P.~R.} \bibnamefont{ten Wolde}}
  \bibnamefont{and} \bibinfo{author}{\bibfnamefont{M.~J.}
  \bibnamefont{Ruiz-Montero}}, \bibinfo{journal}{J.\ Chem.\ Phys.}
  \textbf{\bibinfo{volume}{104}}, \bibinfo{pages}{9932} (\bibinfo{year}{1996}).

\bibitem[{\citenamefont{Radu and Schilling}(2013)}]{radu13}
\bibinfo{author}{\bibfnamefont{M.}~\bibnamefont{Radu}} \bibnamefont{and}
  \bibinfo{author}{\bibfnamefont{T.}~\bibnamefont{Schilling}},
  \bibinfo{journal}{arXiv:1301.5592}  (\bibinfo{year}{2013}).

\end{thebibliography}
\end{document}